\documentstyle[editedvolume,psfig]{crckapb}

\newcommand{\pdrv}[2]{\frac{\partial #1}{\partial #2}}
\newcommand{\drv}[2]{{{{\rm d} #1}\over {{\rm d} #2}}}
\begin{opening}
\title{SPIN-ORBIT COUPLINGS IN X-RAY BINARIES}
\subtitle{Detailed calculations of mass transfer including tidal forces}

\author{T.M. TAURIS \& G.J. SAVONIJE}
\institute{Center for High-Energy Astrophysics, University of Amsterdam}

\end{opening}
\runningtitle{SPIN-ORBIT COUPLINGS IN X-RAY BINARIES}

\begin{document}

\begin{abstract}
We discuss the influence of tidal spin-orbit interactions on the orbital dynamics
of close intermediate-mass {X}-ray binaries.
In particular we consider here a process in which spin angular momentum of
a contracting RLO donor star, in a synchronous orbit, is converted into orbital
angular momentum and thus helps to stabilize the mass transfer by widening
the orbit. Binaries which would otherwise suffer from dynamically unstable
mass transfer (leading to the formation of a common envelope and spiral-in
evolution) are thus shown to survive a phase of extreme mass transfer
on a sub-thermal timescale.
Furthermore, we discuss the orbital evolution prior to RLO in {X}-ray binaries
with low-mass donors, caused by the competing effects of wind mass loss
and tidal effects due to expansion of the (sub)giant. 
\end{abstract}

\section{Introduction}
Tidal torques act to establish synchronization between the spin of the
non-degenerate companion star and the orbital motion. Whenever the spin
angular velocity of the donor is perturbed 
(by a magnetic stellar wind; or change in its moment of inertia due to
either expansion or mass loss in response to RLO) the tidal spin-orbit
coupling will result in a change in the orbital angular momentum
leading to orbital shrinkage or expansion.\\ 
We have performed detailed numerical calculations of the non-conservative
evolution of $\sim 200$ close binary systems with $1.0-5.0\,M_{\odot}$ donor
stars and a $1.3\,M_{\odot}$ accreting neutron star.
Rather than using analytical expressions for simple polytropes, we calculated
the thermal response of the donor star to mass loss, using an updated version
of Eggleton's numerical computer code, in order to
determine the stability and follow the evolution of the mass transfer.
We refer to Tauris \& Savonije (1999) for a more detailed description of the
computer code and the binary interactions considered.

\section{The orbital angular momentum balance equation}
Consider a circular\footnote{This is a good approximation since tidal effects
acting on the near RLO giant star will circularize the orbit
on a short timescale of $\sim\!10^4$ yr, cf. Verbunt \& Phinney (1995).}
binary with an (accreting) neutron star and a companion (donor) star
with mass $M_{\rm NS}$ and $M_2$, respectively.
The orbital angular momentum is given by:
$J_{\rm orb} = (M_{\rm NS}\,M_2\,/M)\,\Omega\,a^2$, where $M=M_{\rm NS}+M_2$
and $\Omega = \sqrt{GM/a^3}$ is the orbital angular velocity.
A simple logarithmic differentiation of this equation yields
the rate of change in orbital separation:
\begin{equation}
  \frac{\dot{a}}{a} = 2\frac{\dot{J}_{\rm orb}}{J_{\rm orb}}
                     -2\frac{\dot{M}_{\rm NS}}{M_{\rm NS}}
                     -2\frac{\dot{M}_2}{M_2}
                     +\frac{\dot{M}_{\rm NS}+\dot{M}_2}{M}
\end{equation}
where the total change in orbital angular momentum can be expressed as:
\begin{equation}
 \frac{\dot{J}_{\rm orb}}{J_{\rm orb}} =
  \frac{\dot{J}_{\rm gwr}}{J_{\rm orb}} + \frac{\dot{J}_{\rm mb}}{J_{\rm orb}}
  +\frac{\dot{J}_{\rm ls}}{J_{\rm orb}} + \frac{\dot{J}_{\rm ml}}{J_{\rm orb}}
\end{equation}
The first term on the right side of this equation governs the loss of 
$J_{\rm orb}$ due to gravitational wave radiation
(Landau \& Lifshitz 1958). The second term arises due to a combination
a magnetic wind of the (low-mass) companion star and a tidal synchronization
(locking) of the orbit. This mechanism of exchanging orbital into spin
angular momentum is referred to as magnetic braking 
(see e.g. Verbunt \& Zwaan 1981; Rappaport et al. 1983).

\subsection{Tidal torque and dissipation rate}
The third term in eq.(2) was recently discussed by Tauris \& Savonije (1999) and
describes possible exchange of angular momentum between the orbit and the
donor star due to its expansion or mass loss (note, we have neglected
the tidal effects on the gas stream and the accretion disk).
For both this term and the magnetic braking term we estimate
whether or not the tidal torque is sufficiently strong to keep the donor star
synchronized with the orbit.
We estimate the tidal torque due to the interaction between the tidally induced
flow and the convective motions in the stellar envelope by means of the simple
mixing-length model for turbulent viscosity $\nu=\alpha H_{\rm p} V_{\rm c}$, where the
mixing-length parameter $\alpha$ is adopted to be 2 or 3, $H_{\rm p}$ is the local
pressure scaleheight, and $V_{\rm c}$ the local characteristic convective velocity.
The rate of tidal energy dissipation can be expressed as (Terquem~et~al. 1998):
\begin{equation}
   \drv{E}{t}=-\frac{192 \pi}{5} \Omega^2 \int_{R_i}^{R_o} \rho r^2
  \nu \left[\left(\pdrv{\xi_r}{r}\right)^2+6 \left(\pdrv{\xi_h}{r}\right)^2
  \right] \, dr
\end{equation}
where the integration is over the convective envelope and $\Omega$ is the
orbital angular velocity, i.e.  we neglect effects of stellar rotation. The
radial and horizontal tidal displacements are approximated here by
the values for the adiabatic equilibrium tide:
\begin{equation}
  \xi_r= f r^2 \rho \left(\drv{P}{r}\right)^{-1} \qquad \qquad
  \xi_h=\frac{1}{6 r} \drv{(r^2 \xi_r)}{r}
\end{equation}
where for the dominant quadrupole tide ($l\!=\!m\!=2$) $f=-GM_2/(4a^3)$.\\
The locally dissipated tidal energy is taken into account as an extra energy
source in the standard energy balance equation of the star, while the
corresponding tidal torque follows as: $ \Gamma = -(1/\Omega)(dE/dt) $.\\
The thus calculated tidal angular momentum exchange $dJ= \Gamma dt$ between the
donor star and the orbit during an evolutionary timestep $dt$ is taken into
account in the angular momentum balance of the system. If the so calculated
angular momentum exchange is larger than the amount required to keep the donor
star synchronous with the orbital motion of the compact star we adopt
a smaller tidal angular momentum exchange (and corresponding tidal dissipation
rate) that keeps the donor star exactly synchronous.

\subsection{Super-Eddington accretion and isotropic re-emission}
The last term in eq.(2) is the most dominant contribution and is caused 
by loss of mass from
the system (see e.g. van~den~Heuvel 1994; Soberman et al. 1997). 
We have adopted the "isotropic re-emission" model in which
all of the matter flows over, in a conservative
way, from the donor star to an accretion disk in the vicinity of the neutron
star, and then a fraction, $\beta$ of this material is ejected isotropically
from the system with the specific orbital angular momentum of the neutron star.
If the mass-transfer rate exceeds the Eddington accretion limit for the neutron star
$\beta >0$. In our calculations we assumed
$\beta = max [0,\;1-\dot{M}_{\rm Edd}/\dot{M}_2 ]$ and
$\dot{M}_{\rm Edd}=1.5\times 10^{-8} M_{\odot}$ yr$^{-1}$.

\section{Evolution neglecting spin-orbit couplings}
Assuming $\dot{J}_{\rm gwr}=\dot{J}_{\rm mb}=\dot{J}_{\rm ls}=0$
and $\dot{J}_{\rm ml}/J_{\rm orb}=\beta\,q^2\,\dot{M}_2/(M_2\,(1+q))$
one obtains easily analytical solutions to eq.(1).
\begin{figure}
  \psfig{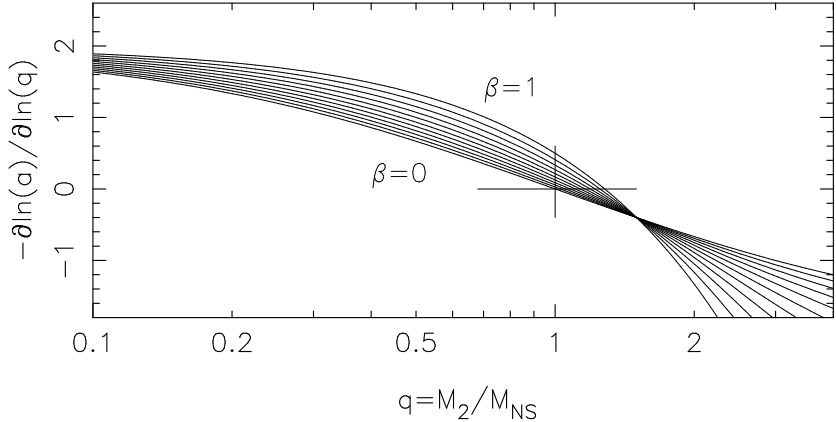}
\caption{
         $-\partial\ln(a)/\partial\ln(q)$ as a function of $q$
         for {X}-ray binaries.
         The different curves correspond to different
         constant values of $\beta$ in steps of 0.1.
         Tidal effects were not taken into account here.
         A cross is shown to highlight
         the case of $q=1$ or $\partial\ln(a)/\partial\ln(q)=0$.
         The evolution during the mass-transfer phase follows
         these curves from right to left since $M_2$ and $q$
         are decreasing with time
         (though $\beta$ need not be constant).}
\end{figure}
In Fig.~1 we have plotted 
\begin{equation}
  -\frac{\partial\ln(a)}{\partial\ln(q)} = 
   2 +\frac{q}{q+1} +q\,\frac{3\beta -5}{q(1-\beta)+1}
\end{equation}
as a function of the mass ratio $q=M_2/M_{\rm NS}$. 
The sign of this quantity is important since
it tells whether the orbit expands or contracts in response to
mass transfer (note $\partial q<0$).
We notice that the orbit always expands when $q<1$
and it always decreases when $q> 1.28$
[solving $\partial\ln(a)/\partial\ln(q)=0$ for $\beta =1$
yields $q=(1+\sqrt{17})/4\approx 1.28$].
If $\beta >0$ the orbit can still expand for $1 < q \le 1.28$.
Note, $\partial\ln(a)/\partial\ln(q)=2/5$ at $q=3/2$ independent of $\beta$.

\section{Results including tidal spin-orbit couplings}
In Fig.~2 we have plotted the orbital evolution of an {X}-ray binary.
The solid lines show the evolution including tidal spin-orbit interactions
and the dashed lines show the calculations without these interactions.
In all cases the orbit will always decrease initially as a result
of the large initial mass ratio ($q=4.0/1.3\simeq 3.1$). But when the
tidal interactions are included the effect of pumping angular momentum
into the orbit (at the expense of spin angular momentum) is clearly seen.
The tidal locking of the orbit acts to convert spin angular
momentum into orbital angular momentum causing the orbit to widen
(or shrink less) in response to mass transfer/loss.
The related so-called Pratt \& Strittmatter (1976) mechanism has previously
been discussed in the literature (e.g. Savonije 1978).
Including spin-orbit interactions many binaries will survive an evolution
which may otherwise end up in an unstable common envelope and spiral-in phase.
An example of this is seen in Fig.~2 where the binary
with initial $P_{\rm orb}=2.5$ days (solid line) only survives as a result of the
spin-orbit couplings. The dashed line terminating at
$M_2 \sim 3.0\,M_{\odot}$ indicates the onset of a run-away mass-transfer process
($\dot{M}_2 > 10^{-3}\,M_{\odot}$ yr$^{-1}$) and formation of a common envelope
and possible collapse of the neutron star into a black hole.
In fact, many of the systems with $2.0 < M_2/M_{\odot} < 5.0$ recently studied by
Tauris, van~den~Heuvel \& Savonije (2000) would not have survived the extreme
mass-transfer phase if the spin-orbit couplings had been neglected.
\begin{figure}
  \psfig{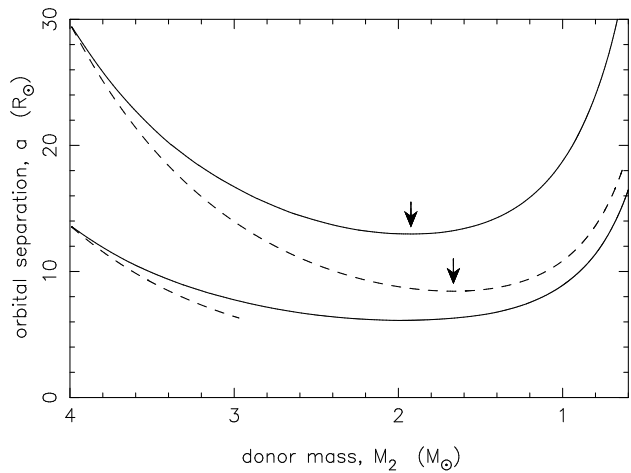}
\caption{Evolution of orbital separation as a function of donor star mass
         during the RLO phase in a binary with $M_2=4.0\,M_{\odot}$
         ({X}=0.70, {Z}=0.02, $\alpha$=2.0),
         $M_{\rm NS}=1.3\,M_{\odot}$ and $P_{\rm orb}=8.0$ and 2.5 days, top and
         bottom lines respectively. The lifetime of these {X}-ray binaries
         are only $t_{\rm X}=1.2$ and 2.1 Myr, respectively.
         The solid evolutionary tracks were calculated including tidal
         interactions and the dashed lines without. See text for details. 
         }
\end{figure}

The location of the minimum orbital separations in Fig.~2 are marked by arrows in
the case of $P_{\rm orb}=8.0$ days. Since the mass-transfer rates in
such an intermediate-mass {X}-ray binary are shown to be highly super-Eddington
(Tauris, van~den~Heuvel \& Savonije 2000) we have $\beta \approx 1$.
Hence in the case of neglecting the tidal interactions (dashed line)
we expect to find the
minimum separation when $q=1.28$ (cf. Section~3).
Since the neutron star at this stage only has accreted 
$\sim 10^{-4}\,M_{\odot}$ we find that the minimum orbital
separation is reached when $M_2=1.28\times 1.30\, M_{\odot}=1.66\,M_{\odot}$.
Including tidal interactions (solid line) results in an earlier spiral-out in the
evolution and the orbit is seen to widen when $M_2 \le 1.92\,M_{\odot}$
($q\approx 1.48$).

\subsection{Low-mass donors and pre-RLO orbital evolution}
For low-mass ($\le 1.5\,M_{\odot}$) donor stars there are two important 
consequences of the spin-orbit interactions which result in a reduction
of the orbital separation: magnetic braking and expansion of the 
(sub)giant companion star. In the latter case the conversion of orbital angular
momentum into spin angular momentum is a caused by a reduced rotation
rate of the donor. 
However, in evolved stars there is a significant wind mass loss 
(Reimers 1975) which will cause the orbit to widen and
hence there is a competition between this effect and the
tidal spin-orbit interactions for determining the orbital evolution
prior to the RLO-phase.
This is demonstrated in Fig.~3.\\
We assumed $\dot{M}_{2\,\rm wind} = -4\times 10^{-13}\;\eta_{\rm RW}\,
L\,R_2/M_2 \;\;\mbox{$M_{\odot}$ yr$^{-1}$}$ where
the mass, radius and luminosity are in solar units and
$\eta_{\rm RW}$ is the mass-loss parameter. 
It is seen that only for binaries with $P_{\rm orb}^{\rm ZAMS} > 100$ days
will the wind mass loss be efficient enough to widen the orbit.
For shorter periods the effects of the spin-orbit interactions dominate
(caused by expansion of the donor) and loss of orbital angular
momentum causes the orbit to shrink.
This result is very important e.g.
for population synthesis studies of the formation of
millisecond pulsars, since $P_{\rm orb}$ in some cases will decrease
significantly prior to RLO.
As an example a system with $M_2 = 1.0\,M_{\odot}$,
$M_{\rm NS}=1.3\,M_{\odot}$ and $P_{\rm orb}^{\rm ZAMS}=3.0$ days
will only have $P_{\rm orb}^{\rm RLO}=1.0$ days
at the onset of the RLO. 

\begin{figure}
  \psfig{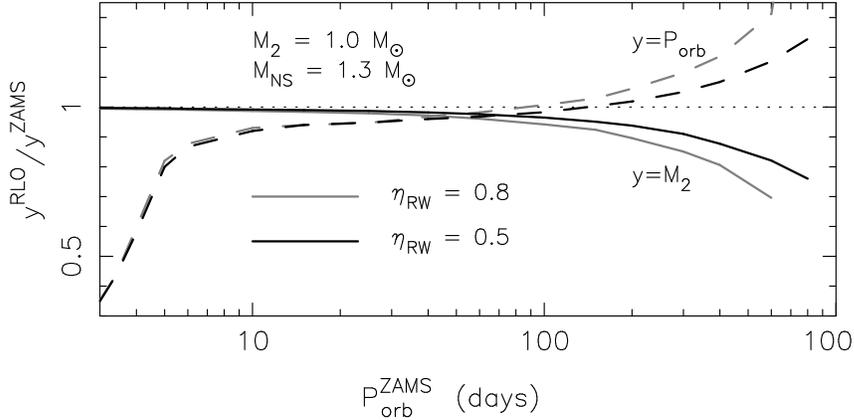}
\caption{The changes of donor mass, $M_2$ (full lines) and orbital
         period, $P_{\rm orb}$ (dashed lines), due to wind mass loss 
         and tidal spin-orbit interactions, from the ZAMS
         until the onset of the RLO as a function of the
         initial orbital period of a circular binary.}
\end{figure}

\end{document}